\documentclass[final]{acmsiggraph}

\usepackage{algorithm}
\usepackage{algpseudocode}
\usepackage[titletoc,title]{appendix}
\usepackage{amssymb}
\usepackage{amsthm}
\usepackage{soul}
\usepackage{amsmath}

\TOGonlineid{99999}
\TOGvolume{0}
\TOGnumber{0}
\TOGarticleDOI{1111111.2222222}
\TOGprojectURL{}
\TOGvideoURL{}
\TOGdataURL{}
\TOGcodeURL{}

\title{Optimisations for Real-Time Volumetric Cloudscapes}

\author{
	Alastair Toft\thanks{e-mail: adt39@cam.ac.uk}
	\and Huw Bowles\thanks{e-mail: huw.bowles@studiogobo.com}\\Studio Gobo
	\and Daniel Zimmermann \thanks{email: daniel.zimmermann@studiogobo.com}\\Studio Gobo}
\pdfauthor{Alastair Toft}

\begin{document}
 \teaser{
   \includegraphics[width=16cm]{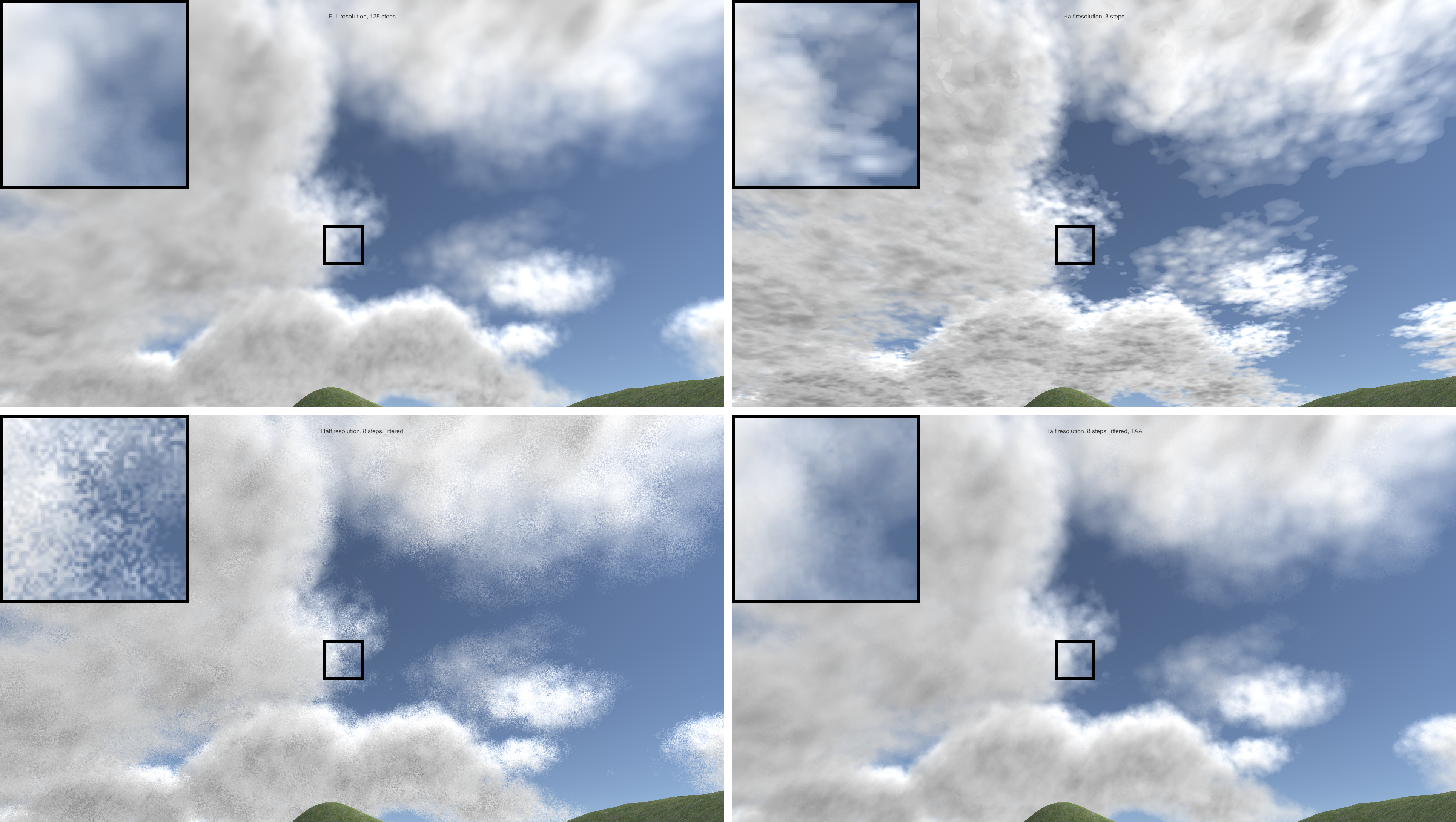}
   \caption{Top left: scene rendered with 128 raymarch steps (draw time 297.7 ms). Top right: rendered with 8 steps, showing much reduced quality (draw time 2.3 ms). Bottom left: rendered with 8 steps with a random offset applied, recovering the appearance of the cloud structure but producing a noisy image (draw time 7.5 ms). Bottom right: as before, with TAA applied to reduce noise (draw time 7.5 ms).}
   \label{fig:teaser}
 }


\maketitle

\begin{abstract}

Volumetric cloudscapes are prohibitively expensive to render in real time without extensive optimisations. Previous approaches render the clouds to an offscreen buffer at one quarter resolution and update a fraction of the pixels per frame, drawing the remaining pixels by temporal reprojection. We present an alternative approach, reducing the number of raymarching steps and adding a randomly jittered offset to the raymarch. We use an analytical integration technique to make results consistent with a lower number of raymarching steps. To remove noise from the resulting image we apply a temporal anti-aliasing implementation. The result is a technique producing visually similar results with $\tfrac{1}{16}$ the number of steps.

\end{abstract}

\section{Introduction}

Our project uses volume raymarching in a pixel shader to draw clouds. Rays are fired into the volume. At each step along the ray, the density is sampled and used to calculate both the pixel's colour and alpha. The raymarch continues until either the alpha reaches 1 or the ray reaches the end of the volume.

To find the alpha of a pixel, we keep a running total of the transmittance along the ray (that is, the fraction of incident light that is absorbed). This is calculated according to Beer-Lambert's law \cite{wrenninge:2011:TR} at each step along the ray.

To calculate the colour of a pixel, we find the incident and ambient lighting at each step. To get incident lighting, we start another ray march from this point on the ray towards the sun (or other light source), and again use Beer-Lambert's law to find the transmittance along this ray. This value is multiplied by the colour and intensity of the light source, as well as a phase function used to approximate the directional scattering that occurs within clouds. We use the Henyey-Greenstein phase function \cite{schneider:2015:RT}. The incident lighting is added to an ambient term which approximates light reflected into the clouds from the ground and the atmosphere.

To get good looking results, this technique requires $\approx$128 raymarch steps with $\approx$6 lighting steps each, per pixel. This makes it far too slow to be used in a game, and extensive optimisations are required to make it close to usable in a real-time situation.

We present a technique to perform the raymarch with an order of magnitude fewer steps.

\section{Previous Work}

The initial, and most significant, influence on this project was the 2015 SIGGRAPH talk by Guerrilla Games \cite{schneider:2015:RT}, which introduced the dynamic volumetric cloudscapes developed for the game Horizon: Zero Dawn. Two main optimisations are described. The first is to perform “cheap” work whenever possible, taking longer steps and only sampling the base shape of the clouds whenever density is lower than a threshold. This method would not be required if fewer raymarch steps were performed in general. The second is to draw the clouds at half of the display resolution, and further to only update a fraction of pixels per frame with new samples, using reprojection to retain pixels from previous frames. This is a fairly complex technique which may introduce visual artifacts as a result of the 4$\times$4 pattern by which pixels are updated.

Playdead \cite{pedersen:2016:TR} introduced their implementation of Temporal Reprojection Anti-Aliasing for the game Inside, which uses information from previous frames to recover subpixel details. This also has the effect of reducing noise present in an image as samples are accumulated over time, allowing for the use of stochastic sampling as described by \cite{gjoel:2016:LC}.

\section{Method}

\subsection{Analytical integration}
\begin{figure}[ht]
\includegraphics[width=\linewidth]{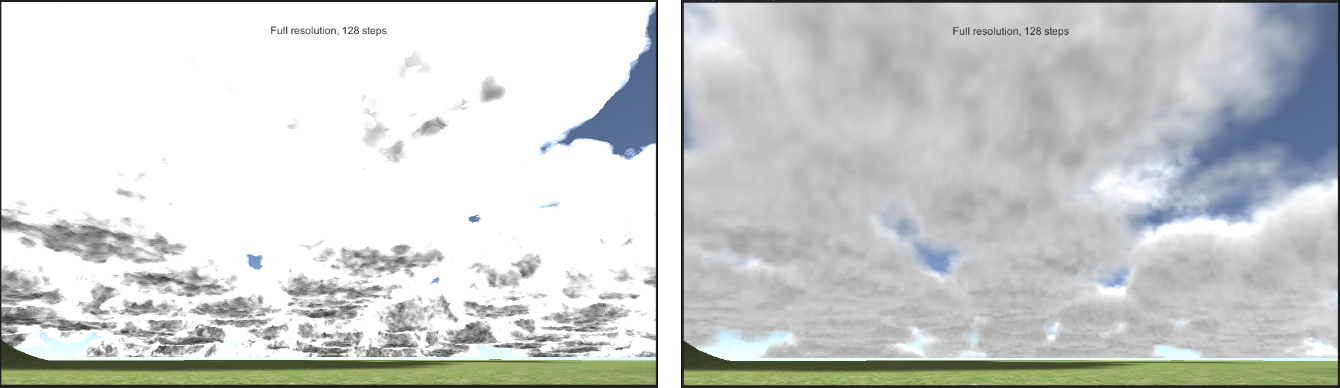}
\caption{Without this change (left), brightness is dependent on step length}
\label{fig:integration}

\end{figure}

We find that, in the standard approach, an inaccuracy causes brightness to be dependent on step length (Figure \ref{fig:integration}). This makes it infeasible to use fewer steps without producing visual artifacts.

The algorithm as described above is, in effect, a discrete approximation of a continuous integration of transmittance with respect to distance through the volume. The basic approach updates the transmittance at a step, and then uses this value to update the colour;
\begin{equation*}
\begin{aligned}
S &= TL, 
\end{aligned}
\end{equation*}
where $S$ is scattering (colour), $T$ is transmittance and $L$ is the lighting term.

This assumes that the transmittance is constant over the step distance, an inaccuracy which leads to the step size dependent output. The solution to this is to analytically integrate the transmittance over the step size, taking only the density to be constant over the step, as follows:

\begin{equation*}
\begin{aligned}
S &= T_0 L \int_{0}^{D} e^{-\rho\alpha x} dx \\
&= T_0 L \left[ \frac{ - e^{-\rho\alpha x} }{\rho \alpha} \right]_0 ^D \\
&= T_0 L \frac{( 1 - e^{-\rho\alpha x})}{\rho \alpha} 
\end{aligned}
\end{equation*}
where $\rho$ is density, $\alpha$ is absorption, $x$ is position, and $D$ is the step distance.

This solves the visual artifacts and allows for larger step sizes to be employed without issue. An analytic integration technique was also mentioned by \cite{hillaire:2016:PB}.
 
\subsection{Temporal anti-aliasing}
TAA has the effect of reducing noise present in an image as information is accumulated from previous frames.

To take advantage of this characteristic, we first significantly reduce the number of raymarch steps taken. This degrades the look of the clouds as much of their structure is lost in between the raymarch samples. To counter this, we apply a random per pixel offset to the start position of the raymarch, such that different depths in the volume are sampled each frame. This recovers most of the visible structure but results in a very noisy image. We apply Playdead's TAA implementation in Unity Engine to the low-res buffer in which the clouds are rendered. After TAA is applied the buffer is composited into the rest of the scene. This reduces this noise to an acceptable level in most cases, with very little performance cost.

\section{Results}

\begin{figure}[ht]
\centering
\begin{tabular}{ |c|c| }
 \hline
 Resolution, render method & Draw time (ms) \\
 \hline \hline
 Full, 128 steps & 297.7 \\
 Half, 128 steps & 128.0 \\
 Half, 8 steps & 2.3 \\
 Half, 8 steps, jitter & 7.5 \\
 Half, 8 steps, jitter, TAA & 7.5 \\
 Quarter, 8 steps, jitter, TAA & 2.4  \\ 
\hline
\end{tabular}
\caption{Draw call timings for a cloud scene running at 1920$\times$1080 on an nVidia GTX 1080. }
\label{fig:timings}

\end{figure}

Figure \ref{fig:teaser} shows the similarity between a cloud scene rendered with the standard number of steps (top left) and using our technique (bottom right). The noise introduced into the image is reduced by TAA to a visually acceptable level in most cases. 

Figure \ref{fig:timings} shows timings for the draw calls in which the cloud shader is computed. These results were computed from the median of five timings in RenderDoc, which have a high variance but give a general impression of relative performance.

We note the increase in draw time when applying the jitter, which we believe to be a result of increased texture cache misses. If an identical random offset is applied to every ray per frame, then the performance hit does not occur, however TAA is unable to improve the appearance of the result in that case.

\section{Conclusion and Future Work}

This technique allows raymarching of a cloud volume with $\tfrac{1}{16}$ the number of steps required by the traditional method. The technique has the potential to be fast enough for rendering real time dynamic cloudscapes in games.

We expect that adapting the TAA filter design with this specific use case in mind could further reduce noise in the final image. In addition, we would like to find a post-process filter to smooth the output without losing the subtler cloud details.

A significant problem with this approach is the loss of cache utilisation caused by introducing the jittered offset, which has a considerable performance impact and could be a potential area for future research.

\bibliographystyle{acmsiggraph}
\bibliography{volsampling}

\begin{thebibliography}{\protect\citename{Gjoel and Svendsen }2016}

\bibitem[\protect\citename{Gjoel and Svendsen }2016]{gjoel:2016:LC}
{\sc Gjoel, M., and Svendsen, M.}, 2016.
\newblock Low complexity, high fidelity - inside rendering.
\newblock GDC 2016, March.

\bibitem[\protect\citename{Hillaire }2016]{hillaire:2016:PB}
{\sc Hillaire, S.}, 2016.
\newblock Physically based sky, atmosphere and cloud rendering in frostbite.
\newblock ACM SIGGRAPH 2016 Physically Based Shading in Theory and Practice,
  July.

\bibitem[\protect\citename{Pedersen }2016]{pedersen:2016:TR}
{\sc Pedersen, L. J.~F.}, 2016.
\newblock Temporal reprojection anti-aliasing in inside.
\newblock GDC 2016, March.

\bibitem[\protect\citename{Schneider and Vos }2015]{schneider:2015:RT}
{\sc Schneider, A., and Vos, N.}, 2015.
\newblock The real-time volumetric cloudscapes of horizon: Zero dawn.
\newblock ACM SIGGRAPH 2015 Advances in Real-Time Rendering, August.

\bibitem[\protect\citename{Wrenninge }2011]{wrenninge:2011:TR}
{\sc Wrenninge, M.}, 2011.
\newblock Production volume rendering fundamentals course notes.
\newblock ACM SIGGRAPH 2011, October.

\end{thebibliography}

\end{document}